# Fabrication of Feedhorn-Coupled Transition Edge Sensor Arrays for Measurement of the Cosmic Microwave Background Polarization

K.L Denis[1], A. Ali[2], J. Appel[2], C.L. Bennett[2], M.P. Chang[1,3], D.T. Chuss[4], F.A. Colazo[1], N. Costen[1,3], T. Essinger-Hileman[2], R. Hu[1,3], T. Marriage[2], K. Rostem[2], K. U-Yen[1], and E.J. Wollack[1]

[1]NASA Goddard Space Flight Center, Greenbelt, MD 20771 USA
[2]Johns Hopkins University, Baltimore, MD 21218 USA
[3]Stinger Ghaffarian Technologies Greenbelt, MD 20770 USA
[4]Villanova University, Villanova, PA 19085 USA

**Abstract** Characterization of the minute cosmic microwave background polarization signature requires multi-frequency, high-throughput precision instrument systems. We have previously described the detector fabrication of a 40 GHz focal plane and now describe the fabrication of detector modules for measurement of the CMB at 90 GHz. The 90 GHz detectors are a scaled version of the 40 GHz architecture where, due to smaller size detectors, we have implemented a modular (wafer level) rather than the chip-level architecture. The new fabrication process utilizes the same design rules with the added challenge of increased wiring density to the 74 TES's as well as a new wafer level hybridization procedure. The hexagonally shaped modules are tile-able, and as such, can be used to form the large focal planes required for a space-based CMB polarimeter. The detectors described here will be deployed in two focal planes with 7 modules each in the Johns Hopkins University led ground-based Cosmology Large Angular Scale Surveyor (CLASS) telescope.

**Keywords** TES • CMB • Wafer Bonding • Micro-Machining • Polarization

K.L. Denis • J. Appel • A. Ali • et. al,

# 1 Introduction

Gravitational waves produced during inflation are predicted to impart a distinct polarization signature on the cosmic microwave background (CMB). Detection of this signature offers an important tool to investigate the high-energy physics of the early Universe. The ground-based Cosmology Large Angular Scale Surveyor (CLASS) [1] is designed to search for this polarized "B-mode" signal in the CMB. CLASS will survey 70% of the sky at four spectral bands (40, 90, 150 and 220 GHz). The detector and focal plane architecture are designed to meet the requirements of high sensitivity and low systematic errors required for this measurement. The sensor architecture [2,3] consists of a broad-band planar orthomode transducer (OMT) that symmetrically couples the independent polarizations in the feedhorn antennas into separate superconducting microstrip with band defining filters. The signal power in the band is then dissipated in a transition edge sensor (TES) bolometer operating at 150 mK.

   A detector module is made up of three components: (1.) a detector wafer consisting of 37 planar superconducting microwave antennas and filters coupled to 74 transition edge sensors, (2.) a micro-machined silicon backshort which also forms a housing providing electromagnetic isolation to the TES, and (3.) a micro-machined photonic choke to improve feedhorn coupling and reduce the required interface flatness. The three components are indium bump bonded to create a detector module. In the following, we give an update on the status of the module fabrication for the 90 GHz focal planes.

# 2 Fabrication

2.1 Orthomode Transducer (OMT) and TES Bolometer Fabrication

 The OMT and TES bolometers form the core of the detector module. We have developed a Silicon-On-Insulator (SOI) based low temperature polymer bonding/sacrificial wafer process that enables the use of single crystal silicon as both low-loss dielectric for superconducting microstrip and as thermal link between the TES membrane and the heat bath with properties that are repeatable and well understood. Compared to the 40 GHz [4] process the 90 GHz process utilizes the same fabrication design rules however we have added a few improvements and these will be highlighted in below. We briefly describe this process here, highlighting improvements that have been made from the 40 GHz fabrication process.

# Fabrication of Feedhorn-Coupled TES Arrays for Measurement of the Cosmic Microwave Background Polarization

Following the 40 GHz fabrication process shown in Fig 1 we start with a SOI wafer with a 5um float zone silicon device layer. The Nb ground plane is patterned on the device layer side. Then the device layer side of the SOI wafer is then polymer bonded with Benzocyclobutene (BCB) to a degeneratively doped ($< 3$ m$\Omega$-cm) silicon wafer. We have reduced the BCB thickness from 2 μm to 1 μm to simplify the required polymer removal later in the process. The original SOI handle wafer is etched away along with the buried silicon oxide. Subsequent fabrication processes are standard with the limitation that the BCB allows fabrication temperatures up to 250 $^{o}$C.

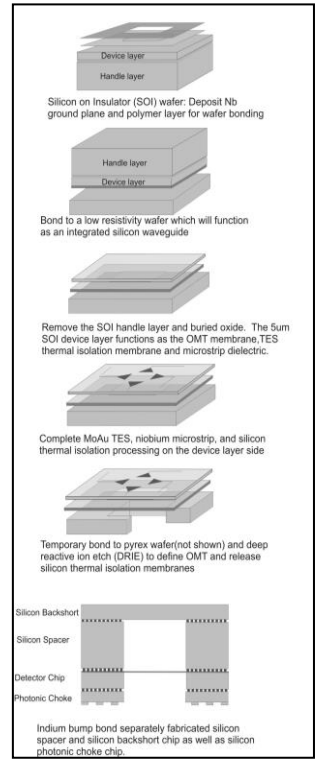

**Fig. 1** Fabrication flow

The next step is TES bilayer deposition. We use MoAu bilayers with a targeted transition temperature ($T_c$) of 150 mK. The MoAu is deposited through a combination of DC magnetron sputtering for the Mo and electron beam evaporation for the Au. We have found through careful control of film thickness uniformity and stress, that using this method we can achieve sufficiently repeatable results [5]. The effects of deposition conditions on Mo Tc have been studied in [6]. We have additionally found that the silicon surface condition can affect the Mo transition temperature by changing the as-deposited Mo film stress for identical deposition conditions. To show this we purposely compared Mo $T_c$ on both hydrophobic and hydrophilic silicon surfaces. Film stress is tuned by controlling the argon sputtering pressure during deposition. All films are 55 nm thick and deposited at 2 A/s with base pressure $< 3 \times 10^{-7}$ torr. Fig 2 shows the variation in Mo $T_c$ as a function of film stress for five separate deposition runs where all surfaces were hydrophobic except in run 4 where an additional separate hydrophilic silicon surface was coated. It can be seen that the hydrophilic film has a $T_c$ 70mK higher than the hydrophobic film. For this reason we are careful to process the TES immediately after the buried oxide removal in HF ensuring a repeatable hydrophobic surface from wafer to wafer.

The niobium microstrip, gold broadband load, PdAu absorber, normal metal bars, and Pd processing as well as the silicon leg etch which incorporates a 35 μm wide, 10 μm long ballistic leg [7] for thermal conductance control all follow the 40 GHz process. The final TES structure is shown in Fig. 2. An additional housing via has been added to reduce out of band leakage. This via (Fig. 3 *right*) is a trench through the silicon dielectric and BCB around the perimeter of the TES and the antenna. It forms a landing where the conductive backshort assembly described below is indium bump bonded directly to the low resistance silicon handle wafer to cut off any spurious leak paths to the TES through the 5 μm thick silicon dielectric. The antenna and TES membrane are formed after deep reactive ion etching (DRIE) through the handle wafer. Photos of the 90 mm diameter hexagonally shaped detector wafer and an individual detector are shown in Fig 3.

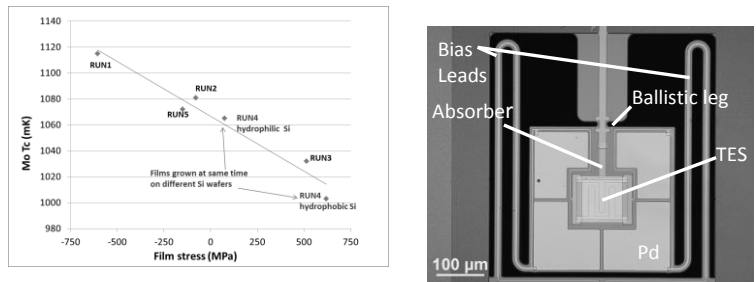

**Fig. 2** (*left*) Variation of molybdenum transition temperature with stress and surface condition. *(right)* Photo of TES and silicon membrane for thermal control.

2.2 Backshort Assembly and Choke Wafer Fabrication

The backshort assembly fabrication follows the process described in [9] scaled to wafer level module sizes. A backshort assembly consists of two degeneratively doped silicon wafers; a spacer wafer and a cap wafer. The spacer wafer sets the approximately quarter wave distance and interfaces with the detector wafer, while the cap provides the reflective backshort. The spacer and cap are bonded by Au-Au thermo-compression bonding. The detector-facing side of the spacer wafer incorporates 10 μm tall indium bumps patterned by liftoff on 50 μm high mesas etched by DRIE around the OMT, TES, and outside perimeter of the module. The mesas

## Fabrication of Feedhorn-Coupled TES Arrays for Measurement of the Cosmic Microwave Background Polarization

incorporate a mouse hole for the niobium microstrip while the indium bumps contact the detector wafer inside the vias described above. This structure provides a micro-machined conductive housing to suppress out of band leakage to the TES. The circular waveguide part of the OMT backshort is formed by a thru-wafer DRIE and is gold coated to reduce microwave loss. A separate silicon choke wafer provides feedhorn coupling [8] to the detector wafer via aluminum coated photonic choke pillars micro-machined by DRIE on the feedhorn side of the wafer. Through wafer etching creates the circular waveguide that is subsequently coated with 1 μm of Al. The choke wafer (Fig. 4 *right*) is bonded by indium bumps patterned on the opposite of the Si pillars. When bonded, the In bumps surround the cavities in the backside of detector wafer around the TES and the circular guide for the Nb antennas.

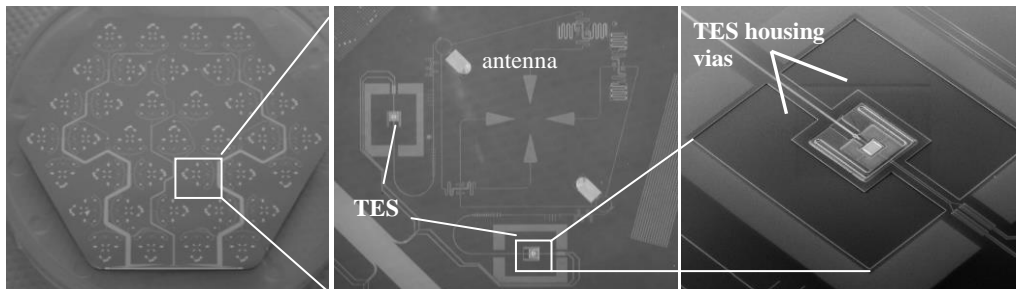

**Fig. 3** (*left*) Photo of detector module. TES wiring is routed to bond pads at bottom edge the hexagon. (*center)* Photo of individual detector. (*right*) SEM of housing vias surrounding TES membrane.

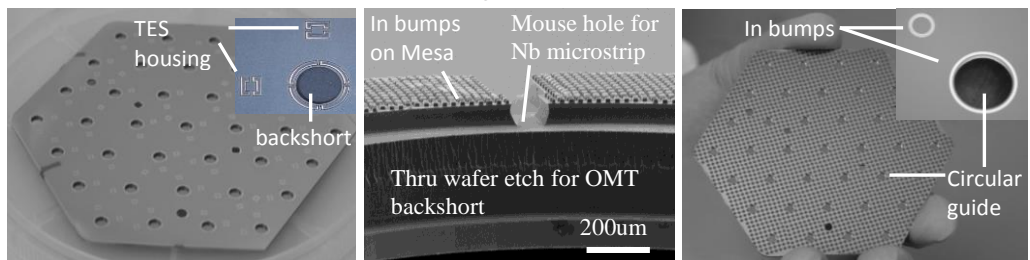

**Fig. 4** (*left*) The backshort assembly is flipped and bonded to the detector wafer front side. (*middle*) Cross-section of mesa and mouse hole along with indium bumps and backshort circular waveguide. (*right*) Photo of photonic choke wafer with inset showing magnified image of indium bump side of wafer which is bonded to the backside of the detector wafer.

2.2 Hybridization

The backshort, detector and choke wafers are hybridized in two steps by flip chip indium bump bonding [10]. The indium bumps are treated to remove the oxide through a proprietary process enabling bonding at room temperature. To ensure a good structural bond the bonding force is set such that the indium is compressed from the initial 10 µm to 5 µm thickness first between the backshort and the detector wafer then between the choke wafer and the backshort/detector wafer assembly. The force applied in the second bond is designed to be half that required in the first simply by scaling the effective indium bump area. Reducing the bond force on the second wafer ensures that there is minimal effect on the alignment of the first bond. The final module consists of a triple stack of choke wafer, detector wafer, and backshort wafer.

## 3 Conclusion

We have described the fabrication of the detector module for the CLASS 90 GHz focal plane. The fabrication process is a scaled version from chip level to wafer level of the currently deployed 40 GHz focal plane utilizing the same fabrication design rules. New features, including integrated TES housing vias, a thinner polymer bonding layer, and micro-machined photonic chokes have been implemented at the wafer level. We have completed fabrication of our first module and it is in thermal and microwave testing at the time of this writing. Fourteen such modules will be incorporated into the two 90 GHz focal planes and deployed at the Atacama Desert in Chile.

**Acknowledgements** NASA ROSES/APRA grant provided support for the detector technology development. We acknowledge the National Science Foundation for their support of CLASS under grants numbered 0959349 and 1429236.

**Fabrication of Feedhorn-Coupled TES Arrays for Measurement of the Cosmic Microwave Background Polarization**